# A HEURISTIC APPROACH FOR WEB-SERVICE DISCOVERY AND SELECTION

Achraf Karray[1], Rym Teyeb[2] and Maher Ben Jemaa[3]

[1] College of Computer Sciences and Information Systems, Najran University, KSA
achraf.karray@gmail.com

[2] Higher Institute of the Technological Studies of Tataouine, Tunisia
rim_teyeb@yahoo.fr

[3] National School of Engineers of Sfax, Tunisia
maher.benjemaa@enis.rnu.tn

## ABSTRACT

*In today's businesses, service-oriented architectures represent the main paradigm for IT infrastructures. Indeed, the emergence of Internet made it possible to set up an exploitable environment to distribute applications on a large scale, and this, by adapting the notion of "service". With the integration of this paradigm in Business to Business Domain (B2B), the number of web services becomes very significant. Due to this increase, the discovery and selection of web services meeting customer requirement become a very difficult operation. Further, QoS properties must be taking into account in the web service selection. Moreover, with the significant number of web service, necessary time for the discovery of a service will be rather long. In this paper, we propose an approach based on a new heuristic method called "Bees Algorithm" inspired from honey bees behavior. We use this technique of optimization in order to discover appropriate web services, meeting customer requirements, in least time and taking into account the QoS properties.*

## KEYWORDS

*Web service discovery and selection; Bees algorithm; peer-to-peer environment; distributed architecture; quality of service (QoS)*

## 1. INTRODUCTION

Web service can be defined as a software entity, independent of platforms, able to be described, published, discovered and composed using the standardized protocols [1, 14, 15, 16]. It is used in various domains such as business to business domain and web-based systems that's why the number of web service's consumer is increasing.

The discovery of web services represents the process allowing the localization of documents describing a Web service (WSDL file [14]). However, with the permanent increase in the number of web services and registries, the operations of discovery and selection become difficult.

The web service discovery and selection are a primordial steps in the service-oriented architectures. In fact, a customer might find several services that provide the same functionality, but can be different in *QoS* properties. Thus, it is essential to adopt an optimization strategy in order to select the best services.

The classical mechanisms of web service discovery are based on the sweeping of all available registries to respond to a client request. With the increase number of services web, these methods are no longer relevant because the discovery time becomes considerable. To get around this problem, innovative techniques of discovery are required. We present in this paper a novel method for discovery and selection of web services based on a genetic algorithm, the bees' algorithm.

This paper is organized as follow. The second section presents related works. In section three, we give an overview of genetic algorithms. Then, we describe the Bees Algorithm in section four. Section five shows the details of the utilization of the Bees Algorithm in the context of web-service discovery and selection. In section six, we present a validation for our approach by implementing it in a P2P (*Peer-to-Peer*) environment. And, we conclude our paper in section seven.

## 2. RELATED WORKS

Discovery is one of the major challenges of the web service technology [22]. Service discovery is the process of finding an appropriate service for a service requestor. The basic steps of this process, shown by figure 1, are:

- Service description: Providers describe their services.
- Service publishing: Services are classified; and their descriptions are published through registries.
- Service discovery: Requestors ask some registries if there are providers offering services with desired capabilities (Description of Requestor's Needs). The subset of the stored descriptions, which match with the customer' request, is returned to the customer (requestor).

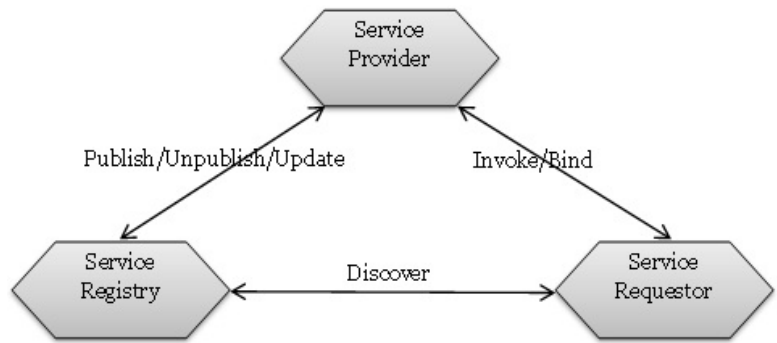

Figure 1. The Web Service Interaction model

The description of web service capabilities is essential for classifying, discovering and using a service. Indeed, service requestors invoke services based upon the discovered service descriptions. Based on the information contained in description files, we distinguish two categories of discovery:

- **Syntactic Discovery:** The discovery process focus on the implementation aspects of a service and thus tailored towards the programmers' requirements. Syntactic discovery can be classified according to the adopted architecture :

  - Centralized architecture: In centralized architecture, description files are kept in the same directory.
  - Distributed (or decentralized) architecture: In this type of architecture, description files are distributed in different servers. Approaches as distributed UDDI [24] and AASDU [25] (*Agent Approach for Service Discovery and Utilization*), adopt this architecture.

- **Semantic discovery:** The discovery process focus on the conceptual aspects of a service aiming to facilitate end-users by shielding off the lower level technical details. Semantic discovery can be classified according to the adopted architecture :

  - Centralized architecture: In centralized architecture, description files are kept in the same directory. Approaches as Ontology Web Language-Service [26] and WSDA [27] (*Web Services Discovery Architecture*) implement this architecture.
  - Distributed (or decentralized) architecture: In this type of architecture, description files are distributed in different servers. Approaches as Speed-R [28] (*Semantic P2P Environment for Diverse web service Registries*), PSWSD [29] (*P2P-based Semantic Web Service Discovery*) and LARKS [23] (*Language for Advertisement and Request for Knowledge Sharing*) adopt a distributed architecture.

The different approaches cited in this section implement a sequential algorithm for set up discovery process. We propose in this paper a method based on a genetic method, the bees algorithm, to find and select appropriate services that matches with the service' requestor. The Application of "Bees Algorithm" in different problems gave excellent results (see section 4.2), so we have had the idea to use this technique of optimization in order to discover and select appropriate web services by taking into account the QoS properties.

## 3. GENETIC ALGORITHMS

### 3.1 Overview

Genetic Algorithms (GA) are part of the Evolutionary Algorithms which are adaptive heuristic methods that exploit ideas of natural selection and biological evolution (such as reproduction, mutation and recombination). In general, evolutionary algorithms implement a same main loop including the following steps (listing 1).

> 1. Initialize and evaluate the initial population.
> 2. Perform competitive selection.
> 3. Apply genetic operators to generate new solutions.
> 4. Evaluate solutions in the population.
> 5. Start again from point 2 and repeat until some convergence criteria is satisfied.

Listing 1. Structure of an evolutionary algorithm

Genetic algorithms find application in bio-informatics, phylogenetic, computational science, engineering, economics, chemistry, manufacturing, mathematics, physics and other fields. There are various flavours of GAs in circulation, varying in implementation of these three parameters, but in essence the algorithms all follow a standard procedure. They all share a common conceptual base of simulating the 'evolution' of individual structures via processes of selection, mutation and reproduction. Once the genetic representation and the fitness function are defined, a GA proceeds to initialize a population of solutions (usually randomly) and then to improve it through repetitive application of the mutation, crossover, inversion and selection operators.

### 3.2 Different steps of a genetic algorithm

Genetic algorithms are essentially useful to resolve search or optimisation problems. In such algorithm, the computer-problem is described using computational models of evolutionary processes. The evolution usually starts from a population of randomly generated individuals and happens in generations. In each generation, the fitness of every individual in the population is evaluated, multiple individuals are stochastically selected from the current population (based on their fitness), and modified (recombined and possibly randomly mutated) to form a new population. The new population is then used in the next iteration of the algorithm. The algorithm terminates when either a maximum number of generations has been produced, or a satisfactory fitness level has been reached for the population. If the algorithm has terminated due to a maximum number of generations, a satisfactory solution may or may not have been reached. Figure 2 shows the simplified flowchart of a Genetic Algorithm.

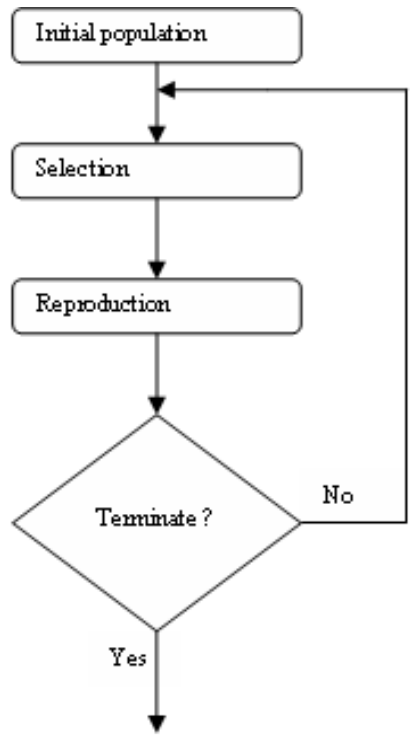

Figure 2. Simplified flow chart of a Genetic Algorithm

A typical genetic algorithm requires a fitness function to evaluate the solution domain. The fitness function is defined over the genetic representation and measures the quality of the represented solution and always depends of the computer-problem.

- **Initialization:** Many individual solutions are (usually) randomly generated to form an initial population. The population size depends on the nature of the problem, but typically contains several hundreds or thousands of possible solutions. Traditionally, the population is generated randomly, allowing the entire range of possible solutions (the *search space*).

- **Selection:** During each successive generation, a proportion of the existing population is selected to breed a new generation. Individual solutions are selected through a fitness-based process, where fitter solutions (as measured by a fitness function) are typically more likely to be selected.

- **Reproduction:** The next step is to generate a second generation population of solutions from those selected through genetic operators: crossover (also called recombination), and/or mutation. These processes ultimately result in the next generation population of solutions that is dif-ferent from the initial generation. Generally the average fitness will have increased by this procedure for the population.

- **Termination:** This generational process is repeated until a termination condition has been reached. Common terminating conditions are :

    - A solution is found that satisfies minimum criteria
    - Fixed number of generations reached
    - Allocated budget (computation time/money) reached
    - The highest ranking solution's fitness is reaching or has reached a plateau such that successive iterations no longer produce better results
    - Manual inspection
    - Combinations of the above

The simple generational genetic algorithm procedure is shown in Listing 2.

> *1- Choose the initial population of individuals*
> *2- Evaluate the fitness of each individual in that   population*
> *3- Repeat on this generation until termination (time limit, sufficient fitness achieved, etc.):*
> *    3.1 Select the best-fit individuals for reproduction*
> *    3.2 Breed new individuals through crossover and mutation operations to give birth to offspring*
> *    3.3 Evaluate the individual fitness of new individuals*
> *    3.4 Replace least-fit population with new individuals*

Listing 2: Pseudo code for genetic Algorithm

## 4. BEES ALGORITHM

When looking for a web service which can be an answer for customer's request, the algorithm of discovery and selection should choose from a wide range of services already developed. To reduce the time needed for Web services' discovery, we will use a new optimization technique called "Bees Algorithm".

The Bees Algorithm represents a new optimization algorithm proposed by Pham and al [2]. It is based on natural foraging behavior of honey bees. The algorithm is considered as one of the most important optimization algorithms due to its successful results in various domains such as clustering problems [3], a resolution of preliminary design problem [4], and artificial neural network training [6].

The foraging behavior of honey bees starts by sending scout bees to search for flower patches. They move randomly from one flower to another. When they return to the hive, those scout bees evaluate the various plots visited and sort them above certain quality threshold which can be measured as a combination of some constituents such as sugar. They choose the (m) best flowers and (e) top-rated patches.

Scout bees associated to select sites perform a dance called "waggle dance". This dance represents a mean of communication in the colony. It provides three types of information: direction of a flower patch, distance from the hive and a quality of nectar. This information helps the colony to send its bees to collect nectar without the use of guide or maps.

More follower bees are sent to more promising patches in order to gather nectar efficiently and quickly.

### 4.1 The basic bees algorithm

The pseudo-code of this algorithm is shown in listing 3.

*1. Initialise population with random solutions.*
*2. Evaluate fitness of the population.*
*3. While (stopping criterion not met)*
*//Forming a new population.*
*4. Select sites for neighborhood search.*
*5. Determine the patch size.*
*6. Recruit bees for selected sites and evaluate fitnesses.*
*7. Select the representative bee from each patch.*
*8. Amend the Pareto optimal set.*
*9. Abandon sites without new information.*
*10. Assign remaining bees to search randomly and evaluate their fitnesses.*
*11. End While.*

Listing 3: Pseudo code for Bees Algorithm

The execution of this algorithm requires a number of parameters (figure 3) to be set namely: number of scoot bees (n), number of patches selected for neighborhood research (m), number of "elite" sites selected out of (m) patches (e), number of bees recruited for (m-e) selected sites (nsp) and stopping criterion.

The algorithm begins with placement of (n) scoot bees, randomly, in the search space.

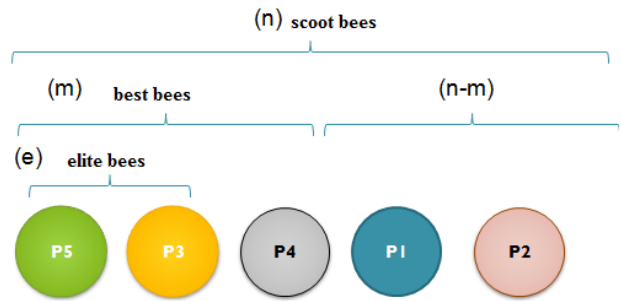

Figure 3. Required parameters for Bees Algorithm

### 4.2 Applications of "Bees Algorithm"

Different approaches use the Bees Algorithm for various problems. In their paper [6], authors propose to apply Bees Algorithm to train RBF networks to control chart pattern recognition. In this work, each bee represented 2345 parameters to be determinate.

In [3], authors use this algorithm to solve clustering problem. In order to partition a data set into homogeneous groups, authors choose to apply the Bees Algorithm which is capable of finding near optimal solution efficiently. A potential clustering solution as set of cluster centers is considered as a bee.

Another application of the bees' algorithm, given in [7], consists to using the Bees Algorithm to optimize parameters of a fuzzy logic controller. It turns parameters of input output membership functions. Therefore each bee was considered as a real number vector representing various parameters to optimize.

Authors of [8] adapted the Bees Algorithm to resolve cell formation problem. Cell formation represents the key problem in designing manufacturing system. Its purpose is to group part with similar processing requirements into part families and associate machines into machine cells. This problem is considered as NP-hard problem because the increase in the problem size causes an exponential increase in the time. In this work, authors use the Bees Algorithm to find the appropriate groups of part families and machine cells which maximize the bond energy metric.

In [9], authors use this algorithm for the Environmental/Economic (Power) Dispatch (EED) problem. In this problem, authors try to minimize, simultaneously, fuel cost and nitrogen oxides emission.

In [2], authors show that the Bees Algorithm is applicable to combinatorial and functional optimization problems. To test the efficacy of this algorithm, they used the inverted Schwefel's function with six dimensions.

# 5. OUR APPROACH

In this paper, we propose a new method which solves the problem of web service discovery. So that, we use an architecture based on registries of web service, called also communities.

Web services' registry  represent a group of web services having the same area of interest and are differentiated for each other according to their non functional properties (such security, time of response…) in order to limit the space of research.  Several approaches used the registry's notion such as in works presented on [5, 10, 11, 12].

The discovery process will be simplified and made more efficient by structuring service registries. To find the optimal service which can be the response of client request, we use the "Bees Algorithm" as an intelligent way of research. Our approach consists of three principals steps:

- Configuration of distributed environment with exploitation of peer-to- peer architecture; every group of peer represents a registry of web services.
- Discovery of the suitable registry having the same business domain with the searched web service. In this step, we use a method based on new population-based optimization algorithm called "Bees algorithm".
- Selection of the optimal web services present in discovered registry. The web service selected has the level of QoS nearest to the value requested by the by the customer.

## 5.1  Architecture

Our architecture consists of:

- A layer for semantic clustering web services in registry. Each registry has its business domain indicated in a description file.
  The update and management of different registries are insured by a control module providing functionality such as adding a new service in the appropriate community.
- A discovery and selection module:  this module ensures discovery of a web service offering the same functionalities searched by customer's request and having the level of quality of service required by the client.

So it is used, in the first place, to find the registry which has the same category with the searched web service, and to choose between different services web present in the registry. The selection is based on level of quality of service.

## 5.2  Adaptation of bees algorithm for web service discovery

The proposed method exploits the capability of the Bees Algorithm to optimize the research procedure. To reduce research area and in order to find a desired web service in the shortest period, we use this technique.  As analogies with the basic Bees Algorithm, we consider distributed registries as a patch and discovery query as a bees.

Figure 4 below summarizes the flow of proposed method. Our proposed algorithm based on "Bees Algorithm" consists of six steps.

- **Step1: random exploration**

  In this step, the algorithm sends discovery queries in parallel way to different registries containing groups of web services. The choice of registries at this stage is randomly.

- **Step2: measure of similarity**

  Step 2 of the algorithm consists to measure the degree of similarity between business domain of wanted web service (searched by a customer) and visited registries. For each registry, a value of fitness function is attributed which indicates the degree of semantic similarity. In a technical way, we use measures of WordNet::Similarity [13] which represents free software that measures the similarity between two terms.

- **Step3 and step4: Sorting and selection of suitable registry**

  - In step 3, the algorithm sorts visited registries according to similarity's degree collected from the second step.
  - In step 4, a registry having the highest value is selected as "elite" registry.

- **Step5: exploration of "elite" registry's neighbourhood**

  In order to explore the neighbourhood of "elite" registry, the algorithm sends research queries to registries resulted from step 4.
  Stopping criterion is to obtain maximal similarity between two areas of interests:  area of searched web service and the selected registry.

- **Step 6: selection of the service providing the required QoS**

  The computation of QoS is done according the formula proposed in [17]. This method determines QoS value based on weights assigned by the service requester. We select the service whose its level QoS corresponds to the value required by the client.

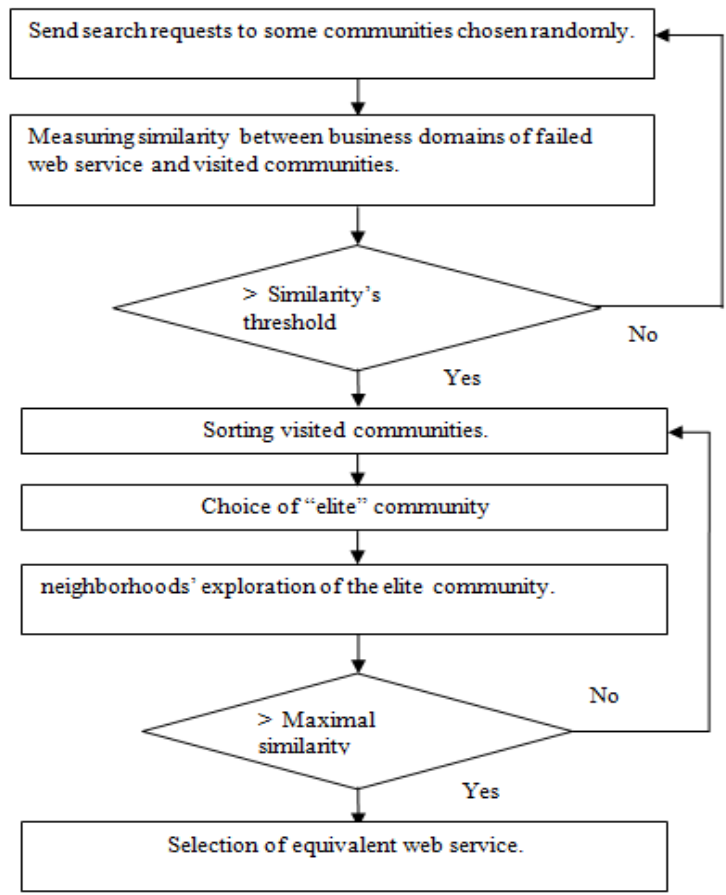

Figure 4. Flowchart for proposed method

# 6. VALIDATION

To validate our approach, we used the JXTA [18] platform, a P2P environment; JDOM [19], an API for XML documents, and WordNet::Similiraty [13] for the similarly between web services. JXTA: it is an open source peer-to-peer protocol specification begun by Sun Microsystems in 2001. This platform allows creating a virtual overlay network. In this virtual network, a given peer could interact with other peers even when some of the peers (or resources) are behind fire-walls (or NATs: Network addresses translation), or use different network transports. In JXTA platform, each resource is identified by a unique ID, so that a peer can change its localization address while keeping a constant identification number.

JXTA platform provides moreover basics services, such as:

- Dynamic search of peers;
- Resources sharing (documents, files, ...) ;
- Creation of groups;
- Secure communication;
- Process Collaboration;
- Private addressing;
- etc …

The JXTA protocols are defined as a set of XML messages which allow any device connected to a network to exchange messages and collaborate independently of the underlying network topology. In fact, JXTA is based upon a set of protocols and standard (such TCP/IP, HTTP, etc.) which are developed in XML language, so that it can be implemented in any programming language and remain independent vis-à-vis any operating system and transport layer protocols. An overview of the JXTA architecture is given in figure 5.

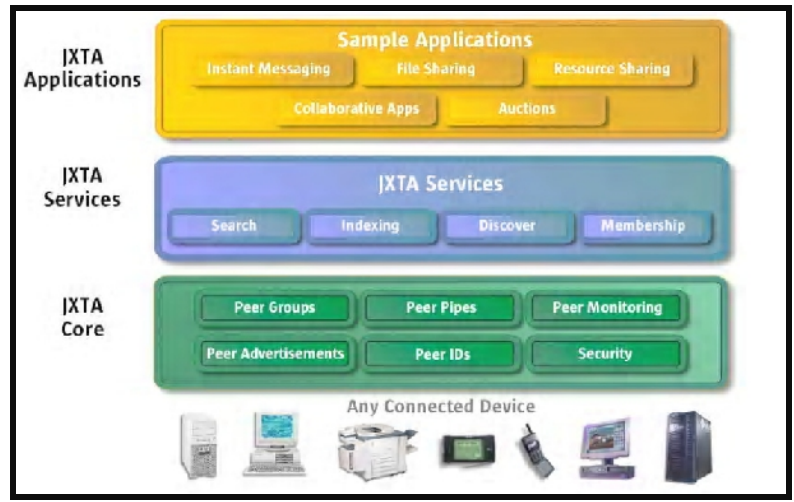

Figure 5. The JXTA architecture

JXTA is an extensible platform which facilitates its adaptation to achieve APIs. This aspect addresses the needs and expectations of developers. JXTA provides some services and notions which are very useful for the implementation of our approach. Among these notions, we find the "announcements" notion that allows us to describe each directory. Moreover, the notion of "peer group", provided by JXTA, is appropriate for our approach to gather directories according their business areas. Also, JXTA incorporates a set of services, as the discovery service, which is based on XML, so that it is easy to adapt these services for our needs.

For all these reasons we opted to use the JXTA platform for implement our approach.

The second tool that we used is JDOM. It is an API provided by W3C allowing modeling and manipulation of XML documents. We use this API essentially to extract, easily, information and data contained in the description files of directory and web-service writing in the XML language.

The last tool which we used is WordNET. We use this tool to find the service that matches the user request. For calculate the degree of similarity between the web-service[1] and the different available directories we use the *WU and Palmer* method [21]. The choice of this method is done after a series of comparison between the different techniques of similarity measure used by WordNet. The advantage of *WU and Palmer* technique it is simpler and more efficient to calculate the similarity degree compared to the other methods.

To validate our approach, we applied the algorithm of discovery and selection (described in section 3) in the case of web service failure. We assume that a given service is failed, and we try to find an equivalent service, in the least delay, while taking in account the properties of QoS (non-functional attributes) such as performance, security, etc. By hypothesis, web services are grouped in different communities. Each community contains equivalent web services (in terms of functional), but that can be differ in their QoS properties.

---

[1] This service can be published by a provider, or requested by a client.

The implemented architecture consists of three main components:

- **Communities layer:** web services are grouped in different communities according to their functionality. Each community is defined by a description file (in XML format), which specifies the business area of the community and the list of web services belonging of this community. A web service is described by its name, its unique identifier, its URL and its QoS attributes.

- **Substitution module:** this module is triggered after a Web service failure. Its function is to find a service providing the same functionalities of the failed service, and having good QoS proprieties. First, the substitution module identifies the community of failed service, and then selects, from the services present in the community, the service providing the QoS required by the client. Finally, the substitution module must add the couple, ***failed service-equivalent service***, in the equivalence cache.

- **Equivalence cache:** it represents a database containing equivalent services. After a service failure, the module of substitution queries the cache to find an entry for the failed service. If this is the case, it extracts and uses the corresponding equivalent service. Thus, the search process can be faster if the couple (***failed web service – equivalent web service***) exists in the cache. Each entry in the cache is valid for a determined period. After expiration of this period, this entry is removed to avoid overloading the database.

Figure 6 describes interactions between different components of our architecture. Once a web service is failed, the substitution module queries the database (cache equivalence) to search an entry for the failed service (step1). Any entry is present in the cache for the queried service (step2). We note here that if there is an entry for the failed service in the database, the next step is not necessary. The substitution module uses the service equivalent to the failed service found in the cache. As the required service is not found, the substitution module selects an equivalent service from the communities (step 3), and finally adds the couple (***failed service-equivalent service***) in the cache (step 4).

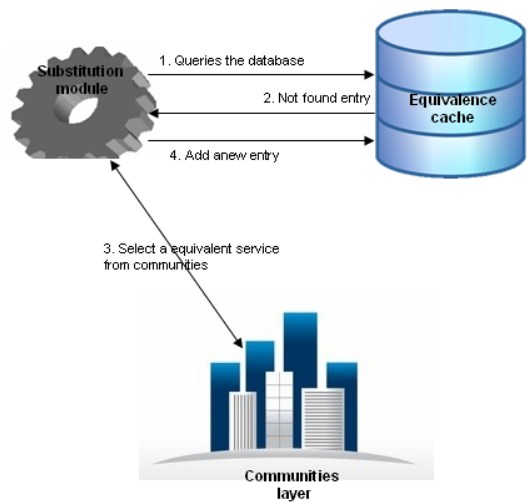

Figure 6. Interactions between components of our architecture

# 7. CONCLUSION

In this paper, we present an innovative method for web service discovery and selection based on heuristic technique called "Bees Algorithm". We validated our approach by implementing it in a P2P environment.

Our approach is based on two important stages. The first consists to find a registry having the same/equivalent business domain of the searched service whilst minimizing the cost of the discovery phase. The second stage consists to calculate the level of QoS of all services present in found registry and select the service having the QoS level nearest to the value requested by the client.